\documentclass[conference]{IEEEtran}
\usepackage{multicol}
\usepackage{etoolbox}
\makeatletter
\patchcmd{\@makecaption}
  {\scshape}
  {}
  {}
  {}
\makeatletter
\patchcmd{\@makecaption}
  {\\}
  {.\ }
  {}
  {}
\makeatother

\usepackage{amsfonts}
\usepackage{amsmath}
\usepackage{mathrsfs}
\usepackage{mathtools}
\usepackage{amsfonts}
\usepackage{amssymb}
\usepackage{graphicx}
\usepackage{epsfig}
\usepackage{psfrag}{}
\usepackage{array}
\usepackage{cases}
\usepackage{eufrak}
\usepackage{cite,graphicx,amssymb,color}
\usepackage{algorithmic}
\usepackage{algorithm}
\usepackage{setspace}
\usepackage{subfigure}
\usepackage{bm}
\usepackage{multirow}
\usepackage{threeparttable}
\usepackage{array}
\usepackage{makecell}
\usepackage{soul}

\newcommand{\tabincell}[2]{\begin{tabular}{@{}#1@{}}#2\end{tabular}}

\newtheorem{Thm}{Theorem}

\newtheorem{Prob}{Problem}

\IEEEoverridecommandlockouts

\begin{document}
\title{Optimal Streaming of 360 VR Videos with Perfect, Imperfect and Unknown FoV Viewing Probabilities}
\author{\IEEEauthorblockN{Lingzhi Zhao, Ying Cui and Chengjun Guo}\IEEEauthorblockA{Shanghai Jiao Tong University, China}\and\IEEEauthorblockN{Zhi Liu}\IEEEauthorblockA{Shizuoka University, Japan}}
\maketitle


\begin{abstract}
In this paper, we investigate wireless streaming of multi-quality tiled 360 virtual reality (VR) videos from a multi-antenna server to multiple single-antenna users in a multi-carrier system. To capture the impact of field-of-view (FoV) prediction, we consider three cases of FoV viewing probability distributions, i.e., perfect, imperfect and unknown FoV viewing probability distributions, and use the average total utility, worst average total utility and worst total utility as the respective performance metrics. We adopt rate splitting with successive decoding for efficient transmission of multiple sets of tiles of different 360 VR videos to their requesting users. In each case, we optimize the encoding rates of the tiles, minimum encoding rates of the FoVs, rates of the common and private messages and transmission beamforming vectors to maximize the total utility. The problems in the three cases are all challenging nonconvex optimization problems. We successfully transform the problem in each case into a difference of convex (DC) programming problem with a differentiable objective function, and obtain a suboptimal solution using concave-convex procedure (CCCP). Finally, numerical results demonstrate the proposed solutions achieve notable gains over existing schemes in all three cases. To the best of our knowledge, this is the first work revealing the impact of FoV prediction and its accuracy on the performance of streaming of multi-quality tiled 360 VR videos.
\end{abstract}


\section{Introduction}
A 360 VR video is generated by capturing a scene of interest in every direction at the same time using omnidirectional cameras. A user wearing a VR headset or head mounted display (HMD) can freely watch the scene of interest in any viewing direction at any time, hence enjoying immersive viewing experience. VR has vast applications in entertainment, education, medicine, etc. A 360 VR video is of a much larger size than a traditional video. At any moment, a user watching a 360 VR video is interested in only one viewpoint, i.e., the center of one part of the 360 VR video, referred to as FoV. To improve transmission efficiency for 360 VR videos, tiling technique is widely adopted. Transmitting only the set of tiles covering the predicted FoV or FoVs with higher viewing probabilities can reduce the required communication resources, while maintaining the quality of experience (QoE) to certain extent. In addition, pre-encoding each tile into multiple representations with different quality levels allows quality adaptation according to user heterogeneity (e.g., in cellular usage costs, display resolutions of devices, channel conditions, etc.). In this paper, we focus on wireless streaming of multi-quality tiled 360 VR videos to multiple users.

Based on viewpoint or FoV prediction, \cite{GLOBALCOM18,TMM20L,JSTSP20,HUANG20191,IS19} study wireless streaming of one multi-quality tiled 360 VR video \cite{GLOBALCOM18,TMM20L,JSTSP20} or multiple multi-quality tiled 360 VR videos \cite{HUANG20191,IS19}. Specifically, it is assumed in  \cite{GLOBALCOM18,TMM20L,JSTSP20,HUANG20191,IS19} that the FoVs that may be watched are known. Furthermore, in \cite{JSTSP20,HUANG20191,IS19}, the viewing probability distributions over those FoVs are also available. Given such FoV prediction results, in \cite{GLOBALCOM18,TMM20L,JSTSP20,HUANG20191,IS19}, the authors optimize the quality level selection and communication resource allocation to maximize the total utility \cite{GLOBALCOM18,HUANG20191,IS19}, minimize the total distortion \cite{JSTSP20}, or minimize the total transmission power \cite{TMM20L}. 

Note that most existing works \cite{GLOBALCOM18,JSTSP20,HUANG20191,TMM20L,IS19} rely on the assumption of perfect FoV prediction. It is unknown how FoV prediction errors influence the performance of wireless streaming of multi-quality tiled 360 VR videos. Besides, note that \cite{GLOBALCOM18,JSTSP20,HUANG20191,TMM20L,IS19} all consider single-antenna servers, which cannot exploit spatial degrees of freedom, and hence cannot provide satisfactory performance for wireless streaming of 360 VR videos. In practice, deploying multiple antennas at a server can significantly improve the performance of wireless systems via designing efficient beamformers, and linear precoders such as zero forcing \cite{TSP08} are widely used. However, any linear precoding scheme can be far from optimal, as the cost to suppress interference can be high when the channels for some of the users are spatially aligned. To circumvent such limitation, the idea of rate splitting which partially decodes interference and partially treats interference as noise \cite{TCOM16} is introduced to improve the overall performance for serving multiple users.

In this paper, we investigate wireless streaming of multi-quality tiled 360 VR videos from a multi-antenna server to multiple single-antenna users in a multi-carrier system. We consider three cases of FoV viewing probability distributions, i.e., perfect, imperfect and unknown FoV viewing probability distributions, and use the average total utility, worst average total utility and worst total utility as the respective performance metrics. We adopt rate splitting with successive decoding for efficient transmission of multiple sets of tiles from different 360 VR videos to their requesting users. In each case of FoV viewing probability distributions, we optimize the encoding rates of the tiles, minimum encoding rates of the FoVs, rates of the common and private messages and beamforming vectors to maximize the total utility. The problems in the three cases are all nonconvex optimization problems. Moreover, the optimization problem in the case of imperfect FoV viewing probability distributions is a max-min problem, and the optimization problem in the case of unknown FoV viewing probability distributions has a non-differentiable objective function. We successfully transform the problem in each case into a DC programming problem with a differentiable objective function, and obtain a suboptimal solution using CCCP\cite{TSP17}. Finally, numerical results show substantial gains of the proposed solutions over existing schemes in all three cases, and reveal the impact of the FoV prediction and its accuracy on streaming of multi-quality tiled 360 VR videos.

\vspace*{-0.15cm}
\section{System Model}\label{section2}
\vspace*{-0.1cm}
\subsection{Multi-Quality Tiled 360 VR Videos}\label{s2_a}
\vspace*{-0.10cm}
We consider tiling to enable flexible transmission of necessary FoVs of a 360 VR video. Specifically, a 360 VR video is divided into $X\times Y$ rectangular segments of the same size, referred to as tiles, where $X$ and $Y$ represent the numbers of segments in each column and each row, respectively. Define $\mathcal{X} \triangleq \{1,\ldots,X\}$ and $\mathcal{Y} \triangleq \{1,\ldots,Y\}$. The ($x,y$)-th tile refers to the tile in the $x$-th row and the $y$-th column, for all $x\in\mathcal{X}$ and $y\in\mathcal{Y}$. For a 360 VR video, consider $\overline{I}$ viewpoints (i.e., $\overline{I}$ FoVs). Denote $\overline{\mathcal{I}} \triangleq\{1,\ldots,\overline{I}\}$. For all $i\in\overline{\mathcal{I}}$, let $\mathcal{F}_{i}$ denote the set of tiles fully or partially included in the $i$-th FoV. Considering user heterogeneity, we pre-encode each tile into $L$ representations corresponding to $L$ quality levels using High Efficiency Video Coding (HEVC), as in Dynamic Adaptive Streaming over HTTP (DASH). Let $\mathcal{L} \triangleq \{1,\ldots, L\}$ denote the set of quality levels. For all $l\in\mathcal{L}$, the $l$-th representation of each tile corresponds to the $l$-th lowest quality. For ease of exposition, assume that the encoding rates of the tiles with the same quality level are identical. The encoding rate of the $l$-th representation of a tile is denoted by $D_{l}$ (in bits/s), where $D_{1} < \ldots < D_{L}$.

\begin{figure}[t]
\begin{center}
 {\resizebox{8.5cm}{!}{\includegraphics{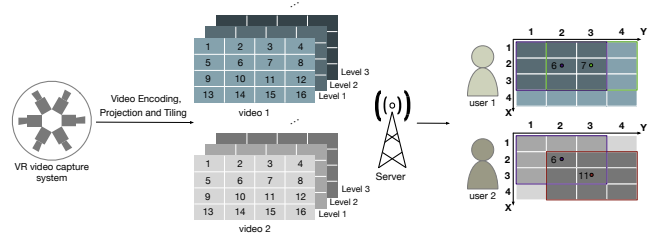}}}
 \vspace*{-0.37cm}
\end{center}
   \caption{\small{System model. $X = 4$, $Y = 4$, $\overline{I} = 16$, $L = 3$, $K = 2$. $\mathcal{I}_{1} = \{6,7\}$, $\mathcal{F}_{6} = \{(1,1),(1,2),(1,3),(2,1),(2,2),(2,3),(3,1),(3,2),(3,3)\}$, $\mathcal{F}_{7} = \{(1,2),(1,3),(1,4),(2,2),(2,3),(2,4),(3,2),(3,3),(3,4)\}$, $r_{6,1} = D_{2}$, $r_{7,1} = D_{1}$, $R_{x,y,1} = D_{2},(x,y)\in\mathcal{F}_{6}$, $R_{x,y,1} = D_{1},(x,y)\in\mathcal{F}_{7}\backslash\mathcal{F}_{6}$. $\mathcal{I}_{2} = \{6,11\}$, $\mathcal{F}_{6} = \{(1,1),(1,2),(1,3),(2,1),(2,2),(2,3),(3,1),(3,2),(3,3)\}$, $\mathcal{F}_{11} = \{(2,2),(2,3),(2,4),(3,2),(3,3),(3,4),(4,2),(4,3),(4,4)\}$, $r_{6,2} = D_{2}$, $r_{11,2} = D_{3}$, $R_{x,y,2} = D_{3},(x,y)\in\mathcal{F}_{11}$, $R_{x,y,2} = D_{2},(x,y)\in\mathcal{F}_{6}\backslash\mathcal{F}_{11}$.}}
   \label{streaming}
\vspace*{-0.60cm}
\end{figure}

As illustrated in Fig. \ref{streaming}, we consider wireless streaming of $K$ multi-quality tiled 360 VR videos from a server to $K$ users, respectively. Let $\mathcal{K}\triangleq \{1,\ldots,K\}$ denote the set of user indices (video indices). We study the duration of the playback time of one group of picture (GOP), which is usually 0.5-1 seconds. A user can freely switch views, when watching a 360 VR video. Suppose the set of FoVs that may be watched by user $k$ and the corresponding FoV viewing probability distribution have been predicted to certain extent. Let $\mathcal{I}_{k}$ represent the sets of indices of the $I_{k}$ FoVs of video $k$ that may be watched by user $k$. Note that for all $k\in\mathcal{K}$, $\mathcal{F}_{i},i\in\mathcal{I}_{k}$ may overlap and user $k$ will watch one of the $I_{k}$ FoVs. Suppose $\mathcal{I}_{k},k\in\mathcal{K}$ are known to the server. For all $k\in\mathcal{K}$ and $i\in\mathcal{I}_{k}$, let $p_{i,k}$ denote the probability that the $i$-th FoV of video $k$ is viewed by user $k$. Here, $p_{i,k} \geq 0, i\in\mathcal{I}_{k},k\in\mathcal{K}$ and $\sum_{i\in\mathcal{I}_{k}}p_{i,k} = 1,k\in\mathcal{K}.$ Denote $\mathbf{p}_{k} \triangleq (p_{i,k})_{i\in\mathcal{I}_{k}},k\in\mathcal{K}$. In the following, we consider three cases of FoV viewing probability distributions.

\noindent\textbf{Perfect FoV viewing probability distributions:} In this case, FoV viewing probability distributions have been estimated by some learning methods, and the estimation errors are negligible. That is, the exact values of $\mathbf{p}_{k},k\in\mathcal{K}$ are assumed to be known to the server.

\noindent\textbf{Imperfect FoV viewing probability distributions:} In this case, FoV viewing probability distributions have been estimated by some learning methods, with certain estimation errors. For all $k\in\mathcal{K}$ and $i\in\mathcal{I}_{k}$, let $\hat{p}_{i,k}$ denote the estimated probability that the $i$-th FoV of video $k$ is viewed by user $k$, and let $\Delta_{i,k} \triangleq p_{i,k} - \hat{p}_{i,k}$ denote the corresponding estimation error, where $\sum\nolimits_{i\in\mathcal{I}_{k}}\Delta_{i,k} = 0$ and $|\Delta_{i,k}| \leq \varepsilon_{i,k}$ for some known $\varepsilon_{i,k} \in (0,1)$. Assume that $\hat{p}_{i,k}$, $\Delta_{i,k},i\in\mathcal{I}_{k},k\in\mathcal{K}$ are known to the server, and $\mathbf{p}_{k},k\in\mathcal{K}$ are not known to the server. Thus, the server knows that the exact FoV viewing probability distributions satisfy $\mathbf{p}_{k}\in\mathcal{P}_{k}$, where
\begin{equation}
\setlength{\abovedisplayskip}{-0.01cm}
\setlength{\belowdisplayskip}{-0.05cm}
\mathcal{P}_{k}\triangleq\left\{\mathbf{p}_{k}~\Big|~\underline{p}_{i,k} \leq p_{i,k} \leq \overline{p}_{i,k},i\in\mathcal{I}_{k},\sum\limits_{i\in\mathcal{I}_{k}}p_{i,k} =1  \right\},k\in\mathcal{K},\nonumber
\end{equation}
with $\underline{p}_{i,k} \triangleq \max\{\hat{p}_{i,k}-\varepsilon_{i,k}, 0\}$ and $\overline{p}_{i,k} \triangleq \min\{\hat{p}_{i,k} + \varepsilon_{i,k}, 1\}$, $i \in \mathcal{I}_{k},k\in\mathcal{K}$.

\noindent\textbf{Unknown FoV viewing probability distributions:} In this case, there is no prior information about the exact FoV viewing probability distributions $\mathbf{p}_{k},k\in\mathcal{K}$.

For all $k\in\mathcal{K}$, the tiles in $\overline{\mathcal{F}}_{k}\triangleq\mathop\cup\nolimits_{i\in\mathcal{I}_{k}}\mathcal{F}_{i}$ may be transmitted to user $k$. Let $R_{x,y,k}$ (in bits/s) denote the encoding rate of the $(x,y)$-th tile of video $k$, where
\begin{equation}
\setlength{\abovedisplayskip}{-0.01cm}
\setlength{\belowdisplayskip}{-0cm}
R_{x,y,k} \in\{0,D_{1},\ldots,D_{L}\},~(x,y)\in\overline{\mathcal{F}}_{k},~k\in\mathcal{K}.\label{tile_rate_max}
\end{equation}
Here, $R_{x,y,k} = 0$ indicates that the $(x,y)$-th tile of video $k$ will not be transmitted to user $k$, and $R_{x,y,k} = D_{l}$ indicates that the $l$-th representation of the $(x,y)$-th tile will be transmitted to user $k$. To avoid degrading QoE, we consider a relative smoothness requirement for quality variation in an FoV \cite{TMM20L}:
\begin{equation}
\setlength{\abovedisplayskip}{-0.01cm}
\setlength{\belowdisplayskip}{-0.05cm}
r_{i,k} \leq R_{x,y,k} \leq r_{i,k} + \delta ,~(x,y)\in\mathcal{F}_{i},~i\in\mathcal{I}_{k},~k\in\mathcal{K},\label{rate_smooth}
\end{equation}
where
\begin{equation}
\setlength{\abovedisplayskip}{-0.01cm}
\setlength{\belowdisplayskip}{-0.05cm}
r_{i,k} \in\{0,D_{1},\ldots,D_{L}\},~i\in\mathcal{I}_{k},~k\in\mathcal{K}.\label{fov_rate_max}
\end{equation}
Here, $r_{i,k}$ (in bits/s) represents the minimum of the encoding rates of the tiles in the $i$-th FoV of video $k$, also referred to as the minimum encoding rate of the $i$-th FoV of video $k$, and indicates the quality level of the $i$-th FoV for user $k$; and $\delta > 0$ is a small number representing the tolerance for quality variation in an FoV (note that the quality variation for tiles in one FoV is not visible if $\delta$ is small enough). An illustration example can be found in Fig. \ref{streaming}.

Let $U(r)$ denote the utility for an FoV with the minimum encoding rate $r$. Here, the utility function $U(\cdot)$ can be any nonnegative, strictly increasing and strictly concave function, and $U(0) = 0$. Let $Q^{(t)}(\mathbf{r}), t =$ pp, ip and up denote the performance metrics in the three cases of FoV viewing probability distributions. In the case of perfect FoV viewing probability distributions (i.e., case-pp), we use the average total utility as the performance metric. In the case of imperfect FoV viewing probability distributions (i.e., case-ip), we use the worst average total utility as the performance metric. In the case of unknown FoV viewing probability distributions (i.e., case-up), we use the worst total utility as the performance metric. Therefore, the performance metrics in the three cases of FoV viewing probability distributions can be written as:
\begin{align}
Q^{(t)}(\mathbf{r}) = \left\{ \begin{array}{ll}
 \sum_{k\in\mathcal{K}}\sum_{i\in\mathcal{I}_{k}}p_{i,k}U(r_{i,k}), & t = \text{pp},\\
 \sum_{k\in\mathcal{K}}\min_{\mathbf{p}_{k}\in\mathcal{P}_{k}} \sum_{i\in\mathcal{I}_{k}}p_{i,k}U(r_{i,k}), & t = \text{ip},\\
 \sum_{k\in\mathcal{K}} \min_{i\in\mathcal{I}_{k}}~U(r_{i,k}), & t = \text{up}.
  \end{array} \right.\label{utility}
\end{align}

\subsection{Physical Layer Model and Transmission Scheme}
The server has $M$ antennas, and each user has one antenna. We consider a multi-carrier system. Let $N$ and $\mathcal{N} \triangleq\{1, \ldots, N\}$ denote the number of subcarriers and the set of subcarrier indices, respectively. The bandwidth of each subcarrier is $B$ (in Hz). We assume block fading, i.e., the channel on each subcarrier remains constant over the considered time duration. Let $\mathbf{h}_{k,n}^{H} \in\mathbb{C}^{1\times M}$ denote the downlink channel vector on subcarrier $n$ between user $k$ and the server. We assume that the channel state information is perfectly known at the server and the users. 

For all $k\in\mathcal{K}$, encoded (source coding) bits of the tiles in $\overline{\mathcal{F}}_{k}$ that will be transmitted to user $k$ are ``aggregated" into one message. We consider a rate splitting scheme \cite{TCOM16}. Specifically, for all $k\in\mathcal{K}$, the aggregated message for user $k$ is further split into a common part of rate $d_{c,k}$ and a private part of rate $d_{p,k}$. Thus, we have:
\begin{equation}
\setlength{\abovedisplayskip}{-0.01cm}
\setlength{\belowdisplayskip}{-0.02cm}
\sum\nolimits_{(x,y)\in\overline{\mathcal{F}}_{k}}R_{x,y,k} = d_{c,k} + d_{p,k},~k\in\mathcal{K}\label{sum_rate_constraint}.
\end{equation} 
Further, the common parts of the messages of the $K$ users are combined into a common message of rate $\sum\nolimits_{k\in\mathcal{K}}d_{c,k}$. The private part of user $k$'s message is also referred to as user $k$'s private message. The common message and the $K$ users' private messages are then encoded (channel coding) into codewords that span over $N$ subcarriers, respectively. Let $\mathbf{w}_{c,n}\in\mathbb{C}^{M\times1}$ and $\mathbf{w}_{k,n}\in\mathbb{C}^{M\times1}$ denote the common beamforming vector and the private beamforming vector for user $k$ on subcarrier $n$, respectively. We have the total transmission power constraint:
\begin{equation}
\setlength{\abovedisplayskip}{-0.01cm}
\setlength{\belowdisplayskip}{-0cm}
\sum\nolimits_{n\in\mathcal{N}}\left(||\mathbf{w}_{c,n}||_{2}^{2} + \sum\nolimits_{k\in\mathcal{K}}||\mathbf{w}_{k,n}||_{2}^{2}\right) \leq P,\label{rs_power_constraint}
\end{equation}
where $P$ is the total transmission power budget. Let $s_{c,n}$ and $s_{k,n}$ denote a symbol of the common message and a symbol of user $k$'s private message, which are transmitted on the $n$-th subcarrier, respectively. For notation simplicity, define $\overline{\mathcal{K}} \triangleq \mathcal{K} \cup \{c\}$. Let $\mathbf{s}_{n} \triangleq (s_{k,n})_{k\in\overline{\mathcal{K}}}$ and assume that $\mathbb{E}[\mathbf{s}_{n}\mathbf{s}^{H}_{n}] = \mathbf{I},$ $n\in\mathcal{N}$. We consider linear precoding on each subcarrier. The received signal at user $k$ on subcarrier $n$ is given by:
\begin{align}
y_{k,n} = ~\mathbf{h}_{k,n}^{H}\mathbf{w}_{c,n}s_{c,n} + &\mathbf{h}_{k,n}^{H}\mathbf{w}_{k,n}s_{k,n} + \sum_{j\in\mathcal{K},j\not=k}\mathbf{h}_{n,j}^{H}\mathbf{w}_{j,n}s_{j,n}\label{received_signal}\nonumber\\
&~~~~~~~+z_{k,n},~k\in\mathcal{K},~n\in\mathcal{N},
\end{align}
where $z_{k,n} \sim \mathcal{CN}(0, \sigma^{2})$ represents the received Additive White Gaussian Noise (AWGN) at user $k$ on subcarrier $n$. 

We consider successive decoding at each user. Specifically, the decoding procedure for user $k\in\mathcal{K}$ is as follows. First, user $k$ decodes the common message by treating the interference from the $K$ users' private messages on each subcarrier as noise. After successfully decoding and removing the common message, user $k$ decodes his private message by treating the interference from the remaining $K-1$ users' private messages on each subcarrier as noise. The Signal to Interference plus Noise Ratios (SINRs) of the common message and user $k$'s private message on subcarrier $n$ are given by $\frac{|\mathbf{h}_{{k,n}}^{H}\mathbf{w}_{c,n}|^{2}}{\sum\nolimits_{j\in\mathcal{K}}|\mathbf{h}_{{k,n}}^{H}\mathbf{w}_{j,n}|^{2} + \sigma^{2}}$ and $\frac{|\mathbf{h}_{{k,n}}^{H}\mathbf{w}_{{k,n}}|^{2}}{\sum\nolimits_{j\in\mathcal{K},j\not=k}|\mathbf{h}_{{k,n}}^{H}\mathbf{w}_{j,n}|^{2} + \sigma^{2}}$, respectively. Assuming Gaussian signaling, 
to guarantee that each user can successfully receive the common message and his private message, we have:
\begin{align}
&\sum\nolimits_{k\in\mathcal{K}}d_{c,k} \leq \sum\nolimits_{n\in\mathcal{N}}B{\rm log}_2\left(1+\frac{|\mathbf{h}_{{k,n}}^{H}\mathbf{w}_{c,n}|^{2}}{\sum\nolimits_{j\in\mathcal{K}}|\mathbf{h}_{{k,n}}^{H}\mathbf{w}_{j,n}|^{2} + \sigma^{2}}\right),\nonumber\\
&~~~~~~~~~~~~~~~~~~~~~~~~~~~~~~~~~~~~~~~~~~~~~~~~~~~~~~~~k\in\mathcal{K},\label{conmon_rate_constraint}
\end{align}
\begin{align}
&d_{p,k} \leq \sum\nolimits_{n\in\mathcal{N}}B{\rm log}_2\left(1+\frac{|\mathbf{h}_{{k,n}}^{H}\mathbf{w}_{{k,n}}|^{2}}{\sum\nolimits_{j\in\mathcal{K},j\not=k}|\mathbf{h}_{{k,n}}^{H}\mathbf{w}_{j,n}|^{2} + \sigma^{2}}\right),\nonumber\\
&~~~~~~~~~~~~~~~~~~~~~~~~~~~~~~~~~~~~~~~~~~~~~~~~~~~~~~~~k\in\mathcal{K}.\label{private_rate_constraint}
\end{align}
\vspace*{-1cm}
\section{Problem Formulation and Optimality Properties}\label{section3}
\vspace*{-0.2cm}
We would like to optimize the encoding rates of the tiles $\mathbf{R} \triangleq (R_{x,y,k})_{(x,y)\in\overline{\mathcal{F}}_{k},k\in\mathcal{K}}$, minimum encoding rates of the FoVs $\mathbf{r} \triangleq (r_{i,k})_{i\in\mathcal{I}_{k},k\in\mathcal{K}}$, rates of the common and private messages $\mathbf{d}\triangleq (d_{c,k},d_{p,k})_{k\in\mathcal{K}}$ and beamforming vectors $\mathbf{w}\triangleq (\mathbf{w}_{k,n})_{k\in\overline{\mathcal{K}},n\in\mathcal{N}}$ to maximize the performance metrics in \eqref{utility} subject to the constraints in \eqref{tile_rate_max}, \eqref{rate_smooth}, \eqref{fov_rate_max}, \eqref{sum_rate_constraint}, \eqref{rs_power_constraint}, \eqref{conmon_rate_constraint}, \eqref{private_rate_constraint}. Note that $\mathbf{R}$ and $\mathbf{r}$ are discrete variables. For tractability, we consider a relaxed version of the discrete optimization, as in \cite{ICC18}. That is, we replace the discrete constraints in \eqref{tile_rate_max} and \eqref{fov_rate_max} with the following continuous constraints:
\begin{align}
&0 \leq R_{x,y,k} \leq D_{L},~(x,y)\in\overline{\mathcal{F}}_{k},~k\in\mathcal{K},\label{multi_tile_relax}\\
&0 \leq r_{i,k} \leq D_{L},~i\in\mathcal{I}_{k},~k\in\mathcal{K}.\label{multi_fov_relax}
\end{align}
Therefore, we formulate the following optimization problem.

\begin{Prob}[Total Utility Maximization]\label{rs_case_general} For $t$ = \text{pp},~\text{ip},~\text{up},
\begin{align}
U^{(t)\star} \triangleq &\max_{\mathbf{R},\mathbf{r},\mathbf{d},\mathbf{w}}\quad Q^{(t)}(\mathbf{r})\nonumber\\
    &\mathrm{s.t.}\quad\eqref{rate_smooth},~\eqref{sum_rate_constraint},~\eqref{rs_power_constraint},~\eqref{conmon_rate_constraint},~\eqref{private_rate_constraint},~\eqref{multi_tile_relax},~\eqref{multi_fov_relax}.\nonumber
\end{align}
Let $(\mathbf{R}^{(t)\star},\mathbf{r}^{(t)\star},\mathbf{d}^{(t)\star},\mathbf{w}^{(t)\star})$ denote an optimal solution of Problem \ref{rs_case_general}.
\end{Prob}

Note that $U^{(t)\star}$ depends on $D_{L}$ and is not related to $D_{1},\ldots,D_{L-1}$. The performance loss induced by the continuous relaxation is acceptable when $D_{2}-D_{1},\dots,D_{L}-D_{L-1}$ are small. Thus, we focus on solving Problem \ref{rs_case_general}. Problem \ref{rs_case_general} is a nonconvex problem, due to the nonconvexity of the constraints in \eqref{conmon_rate_constraint} and \eqref{private_rate_constraint}. There are no effective methods for solving a general nonconvex problem optimally. Besides, Problem \ref{rs_case_general} with $t$ = ip and Problem \ref{rs_case_general} with $t$ = up have non-differentiable objective functions. Therefore, Problem \ref{rs_case_general} is very challenging.\footnote{The proposed method in \cite{TCOM16} for obtaining a suboptimal rate splitting design cannot be used for solving Problem \ref{rs_case_general}.}

Although it is difficult to obtain a globally optimal solution of the nonconvex problem in Problem \ref{rs_case_general}, we can characterize its properties. 
For all $i\in\mathcal{I}_{k},k\in\mathcal{K},$ define $\mathcal{T}_{i,k} \triangleq \mathcal{F}_{i}\backslash\overline{\mathcal{F}}_{k}$. Note that $\mathcal{T}_{i,k} \cap \mathcal{T}_{j,k} = \emptyset,$ for all $i,j\in\mathcal{I}_{k},i\not=j,k\in\mathcal{K}$.
\begin{Thm}[Optimality Properties of Problem \ref{rs_case_general}]\label{theorem}
(\romannumeral1) For $t$ = pp, ip, up, $R^{(t)\star}_{x,y,k} = \max\limits_{i\in\mathcal{I}_{k}: (x,y)\in\mathcal{F}_{i}}r^{(t)\star}_{i,k},(x,y)\in\overline{\mathcal{F}}_{k},~k\in\mathcal{K}$. (\romannumeral2) For $t$ = pp and for all $i,j\in\mathcal{I}_{k},i\not=j,k\in\mathcal{K}$, if $p_{i,k} \leq p_{j,k},$ and $|\mathcal{T}_{i,k}| > |\mathcal{T}_{j,k}| > 1$, then $r^{(\text{pp})\star}_{i,k} \leq r^{(\text{pp})\star}_{j,k}$. For $t$ = ip and for all $i,j\in\mathcal{I}_{k},i\not=j,k\in\mathcal{K}$, if $\overline{p}_{i,k} \leq \underline{p}_{j,k},$ and $|\mathcal{T}_{i,k}| > |\mathcal{T}_{j,k}| > 1$, then $r^{(\text{ip})\star}_{i,k} \leq r^{(\text{ip})\star}_{j,k}$. For $t$ = up, $r^{\text{(up)}\star}_{i,k},i\in\mathcal{I}_{k}$ are identical, $k\in\mathcal{K}$. (\romannumeral3) $U^{(\text{pp})\star} \geq U^{(\text{ip})\star} \geq U^{(\text{up})\star}$.
\end{Thm}

Statement (\romannumeral1) of Theorem \ref{theorem} indicates that in each case, for all $(x,y)\in\overline{\mathcal{F}}_{k},k\in\mathcal{K}$, the first inequality in \eqref{rate_smooth} for at least one FoV that covers the $(x,y)$-th tile is active at an optimal solution. Statement (\romannumeral2) of Theorem \ref{theorem} indicates that in the case of perfect and imperfect FoV viewing probability distributions, an FoV with a higher viewing probability has a higher minimum encoding rate; and in the case of unknown FoV viewing probability distributions, the minimum encoding rates are identical, as all FoVs in $\mathcal{I}_{k}$ are treated the same. Statement (\romannumeral3) shows the relationship among the optimal values of Problem \ref{rs_case_general} for the three cases.

\vspace*{-0.35cm}
\section{Solution}
\vspace*{-0.2cm}
In this section, we obtain a suboptimal solution of Problem \ref{rs_case_general} in each case. First, we transform Problem \ref{rs_case_general} into an equivalent DC programming problem with a differentiable objective function in each case.

\textbf{Perfect FoV viewing probability distributions:} By introducing auxiliary variables and extra constraints, we can equivalently transform Problem \ref{rs_case_general} with $t$ = pp into:
\begin{Prob}[Equivalent Problem of Problem \ref{rs_case_general} with t = pp]\label{rs_case_general_equal}
\begin{align}
&\max_{\mathbf{R},\mathbf{r},\mathbf{d},\mathbf{e},\mathbf{u},\mathbf{w}}\quad Q^{\text{(pp)}}(\mathbf{r})\nonumber\\
    &\mathrm{s.t.}\quad\eqref{rate_smooth},~\eqref{sum_rate_constraint},~\eqref{rs_power_constraint},~\eqref{multi_tile_relax},~\eqref{multi_fov_relax},\nonumber\\
    &\sum\nolimits_{k\in\mathcal{K}}d_{c,k} \leq \sum\nolimits_{n\in\mathcal{N}}e_{c,n},\label{common_sum_rate}\\
    &d_{p,k} \leq \sum\nolimits_{n\in\mathcal{N}}e_{k,n},~k\in\mathcal{K},\label{private_sum_rate}\\
    &\sum\nolimits_{j\in\mathcal{K}}|\mathbf{h}_{{k,n}}^{H}\mathbf{w}_{j,n}|^{2} + \sigma^{2} - \frac{\sum\nolimits_{k\in\overline{\mathcal{K}}}|\mathbf{h}_{{k,n}}^{H}\mathbf{w}_{k,n}|^{2} + \sigma^{2}}{u_{c,n}} \leq 0,\label{dc_common}\nonumber\\
    &~~~~~~~~~~~~~~~~~~~~~~~~~~~~~~~~~~~~~~~~~~~~~~k\in\mathcal{K},~n\in\mathcal{N},
\end{align}
\end{Prob}
\begin{align}
	&\sum\nolimits_{j\in\mathcal{K},j\not=k}|\mathbf{h}_{{k,n}}^{H}\mathbf{w}_{j,n}|^{2} + \sigma^{2} - \frac{\sum\nolimits_{j\in\mathcal{K}}|\mathbf{h}_{{k,n}}^{H}\mathbf{w}_{j,n}|^{2} + \sigma^{2}}{u_{{k,n}}} \leq 0,\label{dc_private}\nonumber\\
  &~~~~~~~~~~~~~~~~~~~~~~~~~~~~~~~~~~~~~~~~~~~~~~k\in\mathcal{K},~n\in\mathcal{N},\\
  &2^{\frac{e_{k,n}}{B}} \leq u_{{k,n}},~k\in\overline{\mathcal{K}},~n\in\mathcal{N},\label{dc_variable_private}
\end{align}
where $\mathbf{e} \triangleq (e_{k,n})_{k\in\overline{\mathcal{K}},n\in\mathcal{N}}, \mathbf{u} \triangleq (u_{k,n})_{k\in\overline{\mathcal{K}},n\in\mathcal{N}}$. Let $(\mathbf{R}^{\text{(pp)}\dag},\mathbf{r}^{\text{(pp)}\dag},\mathbf{d}^{\text{(pp)}\dag},\mathbf{e}^{\text{(pp)}\dag},\mathbf{u}^{\text{(pp)}\dag},\mathbf{w}^{\text{(pp)}\dag})$ denote an optimal solution of Problem \ref{rs_case_general_equal}.
\begin{Thm}[Equivalence between Problem \ref{rs_case_general} with t = pp and Problem \ref{rs_case_general_equal}]\label{lemma_rs_case1}$(\mathbf{R}^{(\text{pp})\dag},\mathbf{r}^{(\text{pp})\dag},\mathbf{d}^{(\text{pp})\dag},\mathbf{w}^{(\text{pp})\dag})$ is an optimal solution of Problem \ref{rs_case_general} with $t$ = pp.
\end{Thm}


\textbf{Imperfect FoV viewing probability distributions:}
$Q^{(\text{ip})}(\mathbf{r})$ is non-differentiable, and Problem \ref{rs_case_general} with $t$ = ip is a max-min problem w.r.t. $(\mathbf{R},\mathbf{r},\mathbf{d},\mathbf{w},\mathbf{p})$. Note that max-min problems are in general very challenging. First, we replace the inner problem with its dual problem which has a differentiable objective function. Next, we introduce the auxiliary variables and extra constraints. Thus, we can equivalently convert Problem \ref{rs_case_general} with $t$ = ip to:

\begin{Prob}[Equivalent Problem of Problem \ref{rs_case_general} with t = ip]\label{rs_case_two_equal}
\begin{align}
&\max_{\substack{\mathbf{R},\mathbf{r},\mathbf{d},\mathbf{e},\mathbf{u},\mathbf{w},\\\bm{\lambda} \succeq 0, \bm{\tau} \succeq 0, \bm{\gamma}}}\sum\limits_{k\in\mathcal{K}}\left(\sum\limits_{i\in\mathcal{I}_{k}}(\tau_{i,k}\underline{p}_{i,k} - \lambda_{i,k}\overline{p}_{i,k}) - \gamma_{k}\right) \nonumber\\
    &\mathrm{s.t.}\quad\eqref{rate_smooth},~\eqref{sum_rate_constraint},~\eqref{rs_power_constraint},~\eqref{multi_tile_relax},~\eqref{multi_fov_relax},~\eqref{common_sum_rate},~\eqref{private_sum_rate},~\eqref{dc_common},~\eqref{dc_private},~\eqref{dc_variable_private},\nonumber\\
    &\quad\quad~U(r_{i,k}) + \lambda_{i,k} - \tau_{i,k} + \gamma_{k} \geq 0, ~i\in\mathcal{I}_{k},~k\in\mathcal{K},\nonumber
\end{align}
where $\bm{\lambda}\triangleq (\lambda_{i,k})_{i\in\mathcal{I}_{k},k\in\mathcal{K}}$, $\bm{\tau}\triangleq (\tau_{i,k})_{i\in\mathcal{I}_{k},k\in\mathcal{K}}$, $\bm{\gamma}\triangleq (\gamma_{k})_{k\in\mathcal{K}}$. Let $(\mathbf{R}^{(\text{ip})\dag},\mathbf{r}^{(\text{ip})\dag},\mathbf{d}^{(\text{ip})\dag},\mathbf{e}^{(\text{ip})\dag},\mathbf{u}^{(\text{ip})\dag},\mathbf{w}^{(\text{ip})\dag},\bm{\lambda}^{(\text{ip})\dag},\bm{\tau}^{(\text{ip})\dag},\\\bm{\gamma}^{(\text{ip})\dag})$ denote an optimal solution of Problem \ref{rs_case_two_equal}.
\end{Prob}

Note that $\lambda_{i,k}$, $\tau_{i,k}$ and $\gamma_{i,k}$ are dual variables of the inner problem, corresponding to $p_{i,k} \leq \overline{p}_{i,k}$, $p_{i,k} \geq \underline{p}_{i,k}$ and $\sum_{i\in\mathcal{I}_{k}}p_{i,k} = 1$, respectively.
\begin{Thm}[Equivalence between Problem \ref{rs_case_general} with t = ip and Problem \ref{rs_case_two_equal}]\label{lemma_rs_case2}$(\mathbf{R}^{(\text{ip})\dag},\mathbf{r}^{(\text{ip})\dag},\mathbf{d}^{(\text{ip})\dag},\mathbf{w}^{(\text{ip})\dag})$ is an optimal solution of Problem \ref{rs_case_general} with $t$ = ip.
\end{Thm}

\textbf{Unknown FoV viewing probability distributions:}
$Q_{\text{up}}(\mathbf{r})$ is non-differentiable. First, we cast Problem \ref{rs_case_general} with $t$ = up in hypograph form which has a differentiable objective function. Next, we introduce the auxiliary variables and extra constraints. Thus, we can equivalently convert Problem \ref{rs_case_general} with $t$ = up to:
\begin{Prob}[Equivalent Problem of Problem \ref{rs_case_general} with t = up]\label{rs_case_three_equal}
\begin{align}
&\max_{\mathbf{R},\mathbf{r},\mathbf{d},\mathbf{e},\mathbf{u}, \mathbf{y} \succeq 0,\mathbf{w}}\sum_{k\in\mathcal{K}}y_{k}\nonumber\\
    \mathrm{s.t.}\quad&\eqref{rate_smooth},~\eqref{sum_rate_constraint},~\eqref{rs_power_constraint},~\eqref{multi_tile_relax},~\eqref{multi_fov_relax},~\eqref{common_sum_rate},~\eqref{private_sum_rate},~\eqref{dc_common},~\eqref{dc_private},~\eqref{dc_variable_private},\nonumber\\
    &y_{k} \leq U(r_{i,k}), ~i\in\mathcal{I}_{k},~k\in\mathcal{K},\nonumber
\end{align}
where $\mathbf{y} \triangleq (y_{k})_{k\in\mathcal{K}}$. Let $(\mathbf{R}^{\text{(up)}\dag},\mathbf{r}^{\text{(up)}\dag},\mathbf{d}^{\text{(up)}\dag},\mathbf{e}^{\text{(up)}\dag},\mathbf{u}^{\text{(up)}\dag},\\\mathbf{y}^{\text{(up)}\dag},\mathbf{w}^{\text{(up)}\dag})$ denote an optimal solution of Problem \ref{rs_case_three_equal}.
\end{Prob}
\begin{Thm}[Equivalence between Problem \ref{rs_case_general} with t = up and Problem \ref{rs_case_three_equal}]\label{lemma_rs_case3}$(\mathbf{R}^{(\text{up})\dag},\mathbf{r}^{(\text{up})\dag},\mathbf{d}^{(\text{up})\dag},\mathbf{w}^{(\text{up})\dag})$ is an optimal solution of Problem \ref{rs_case_general} with $t$ = up.
\end{Thm}

Note that in Problem \ref{rs_case_general_equal}, Problem \ref{rs_case_two_equal} and Problem \ref{rs_case_three_equal}, $\mathbf{u},\mathbf{e}$ are auxiliary variables, and \eqref{common_sum_rate}, \eqref{private_sum_rate}, \eqref{dc_common}, \eqref{dc_private}, \eqref{dc_variable_private} are extra constraints. Furthermore, notice that the constraints in \eqref{common_sum_rate} and \eqref{private_sum_rate} are convex with respect to (w.r.t) $(\mathbf{d},\mathbf{e})$, the constraints in \eqref{dc_variable_private} are convex w.r.t $(\mathbf{e},\mathbf{u})$, and each constraint function in (\ref{dc_common}) and (\ref{dc_private}) can be regarded as a difference of two convex functions w.r.t. $(\mathbf{u},\mathbf{w})$. Therefore, Problem \ref{rs_case_general_equal}, Problem \ref{rs_case_two_equal} and Problem \ref{rs_case_three_equal} are DC programming problems with differentiable objective functions, and their stationary points can be obtained by CCCP\cite{TSP17}. The main idea is to solve a sequence of successively refined approximate convex problems, each of which is obtained by linearizing the second convex part in \eqref{dc_common}, \eqref{dc_private} and preserving the remaining convexity. For each problem, we can run CCCP multiple times with different feasible initial points to obtain multiple stationary points, and choose the stationary point with the best objective value as a suboptimal solution. 

\vspace*{-0.20cm}
\section{Numerical Results}
\begin{figure}[t]
\begin{center}
 {\resizebox{6cm}{!}{\includegraphics{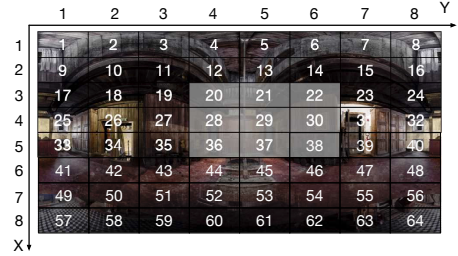}}}
 \vspace*{-0.37cm}
\end{center}
   \caption{\small{Illustration of tiles, viewpoints and FoVs in a 360 VR video. The grey area represents FoV 29 which is centered at viewpoint 29.}}
   \label{example}
\vspace*{-0.60cm}
\end{figure}

In the simulation, we consider the streaming of five 360 VR video sequences, i.e., \textit{Diving}, \textit{Rollercoaster}, \textit{Timelapse}, \textit{Venice}, \textit{Paris}, provided by \cite{inproceedings}. They are indexed by 1, 2, 3, 4 and 5, respectively. As illustrated in Fig. \ref{example}, we divide each 360 VR video into $8 \times 8$ tiles, i.e., $X = 8$, $Y = 8$, and choose $\overline{I} = 64$ FoVs, each of size $3\times 3$ (in the number of tiles). We use \textit{Kvazaar} as the 360 VR video encoder, and set $D_{l},l\in\mathcal{L}$ according to Table \ref{table2}. For each video sequence, based on the viewpoint data of 59 users provided by \cite{inproceedings}, we obtain a viewpoint sequence for each user, with one viewpoint for every 1/3 second. We view users 2, 8, 24, 32 and 40 as the users who request videos 1, 2, 3, 4 and 5, i.e., users 1, 2, 3, 4 and 5, respectively. For video $k$, the viewpoint sequences from the other users besides user $k$ are used for FoV prediction of user $k$. For user $k$, consider the prediction of the 3-rd element in his viewpoint sequence given the 2-nd element. Let $i_{k}$ denote the index of the FoV of user $k$ corresponding to the 2-nd element in his viewpoint sequence. Set $\mathcal{I}_{k} = \{i_{k}-8,i_{k}-1,i_{k},i_{k}+1,i_{k}+8\}$, which contains $i_{k}$ and the indices of the neighbouring FoVs of FoV $i_{k}$. Let $n_{i_{k},i}$ denote the number of users with the 2-nd element and 3-rd element in his viewpoint sequence for video $k$ being $i_{k}$ and $i$, respectively, where $i\in\mathcal{I}_{k}$. Then, we calculate the FoV viewing probabilities according to $p_{i,k} = \frac{n_{i_{k},i}}{\sum\nolimits_{i\in\mathcal{I}_{k}}n_{i_{k},i}},~i\in\mathcal{I}_{k},k\in\mathcal{K}.$ The values of $p_{i,k},i\in\mathcal{I}_{k},k\in\mathcal{K}$ are given in Table \ref{table}. We set $\hat{p}_{i,k} = p_{i,k},\varepsilon_{i,k} =  \varepsilon, i\in\mathcal{I}_{k},k\in\mathcal{K}$, $B$ = 1MHz, $N$ = 128, $P$ = 30 dBm and $\sigma^{2} = 10^{-9}$ W. Besides, we choose $\mathbf{h}_{{k,n}},k\in\mathcal{K},n\in\mathcal{N}$ randomly and independently according to $\mathcal{CN}(0,\mathbf{1}_{M\times M})$, and set the pass loss $\beta_{k} = 1, k \in \mathcal{K}$. We evaluate the average performance metrics over 100 random realizations of $\mathbf{h}_{{k,n}},k\in\mathcal{K},n\in\mathcal{N}$. We adopt the utility function presented in \cite{Zhang2013QoE}, i.e., $U(r) = 0.6 \log(1000\frac{r}{D_{L}})$. For ease of presentation, in the following, the performance metrics in all three cases are referred to as total utility, unless otherwise specified. In case-$t$, the proposed solution of Problem \ref{rs_case_general} is referred to as Prop-$t$, and the feasible solution of the original discrete problem, which is constructed based on Prop-$t$ (as illustrated in Section \ref{section3}), is referred to as Prop-Disc-$t$, where $t$ = pp, ip, up. 

\begin{table}  
\caption{\small{Encoding rates (in Mbit/s) for $L = 3,5,7$.}}
\resizebox{9cm}{!}{  
\begin{tabular}{|c|c|} 
\hline  
$L$& $D_{l}$, $l\in\mathcal{L}.$ \\ \hline  
3 & $D_{1} = 14.46,~D_{2} = 52.97,~D_{3} = 87.75.$   \\ \hline 
5 & $D_{1} = 14.46,~D_{2} = 37.10,~D_{3} = 52.97,~D_{4} = 69.53,~D_{5} = 87.75.$\\ \hline
7 &$D_{1} = 14.46,~D_{2} = 37.10,~D_{3} = 46.20,~D_{4} = 52.97,~D_{5} = 59.45,~D_{6} = 69.53,~D_{7} = 87.75.$\\ \hline
\end{tabular}
} \label{table2}
\end{table}

\begin{table}[t]\large
\caption{\small{Prediction parameters.}}
\begin{center}
\resizebox{8.5cm}{!}{  
\begin{tabular}{|p{1cm}|p{2.2cm}|p{1.7cm}|p{1.7cm}|p{4.2cm}|p{7cm}|} 
\hline  
$k$& Video sequence& User&Current FoV& Predicted FoVs& FoV viewing probability distributions  \\ \hline  
1 & \textit{Diving}  &2& 28 &$\mathcal{I}_{1} = \{20,27,28,29,36\}$& \tabincell{l}{ $(p_{20,1},p_{27,1},p_{28,1},p_{29,1},p_{36,1})$ \\$= (0.4138,0.1724,0.2414,0.1667,0.0417)$ }  \\ \hline  
2 & \textit{Rollercoaster} &8& 21 &$\mathcal{I}_{2} = \{13,20,21,22,29\}$&  \tabincell{l}{$(p_{13,2},p_{20,2},p_{21,2},p_{22,2},p_{29,2})$\\$ = (0,0.4615,0.3077,0.0769,0.1538)$}  \\ \hline  
3 & \textit{Timelapse} &24& 24 &  $\mathcal{I}_{3} = \{16,23,24,17,32\}$&\tabincell{l}{$(p_{16,3},p_{23,3},p_{24,3},p_{17,3},p_{32,3})$\\$ = (0.1481,0.037,0.2963,0.5185,0)$}\\ \hline
4 & \textit{Venice} & 32&29 &$\mathcal{I}_{4} = \{21,28,29,30,37\}$&\tabincell{l}{$(p_{21,4},p_{28,4},p_{29,4},p_{30,4},p_{37,4}) $\\$= (0.25,0.375,0.25,0.0625,0.0625)$} \\ \hline
5 & \textit{Paris} & 40&18 & $\mathcal{I}_{5} = \{10,17,18,19,26\}$&\tabincell{l}{$(p_{10,5},p_{17,5},p_{18,5},p_{19,5},p_{26,5}) $\\$= (0.375,0.5,0.125,0,0)$} \\ \hline
\end{tabular}
} \label{table}
\end{center}
\end{table}

\begin{figure}[t]
\begin{center}
 \subfigure[\small{Total utility versus $L$ at $\varepsilon = 0.4$}]
 {\resizebox{4.2cm}{!}{\includegraphics{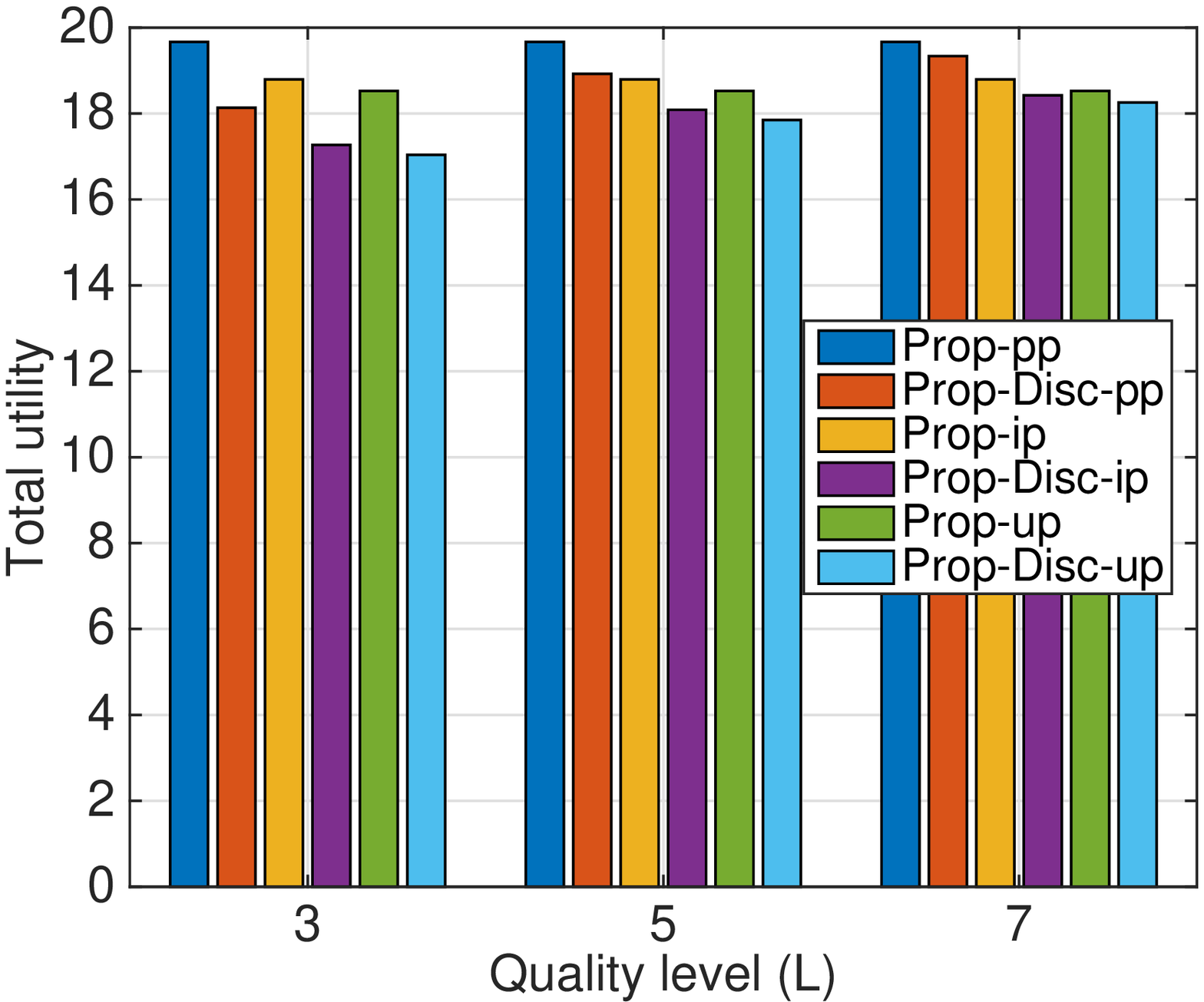}}}
 \vspace*{-0.07cm}
  \subfigure[\small{Worst average total utility versus $\varepsilon$.}]
 {\resizebox{4.2cm}{!}{\includegraphics{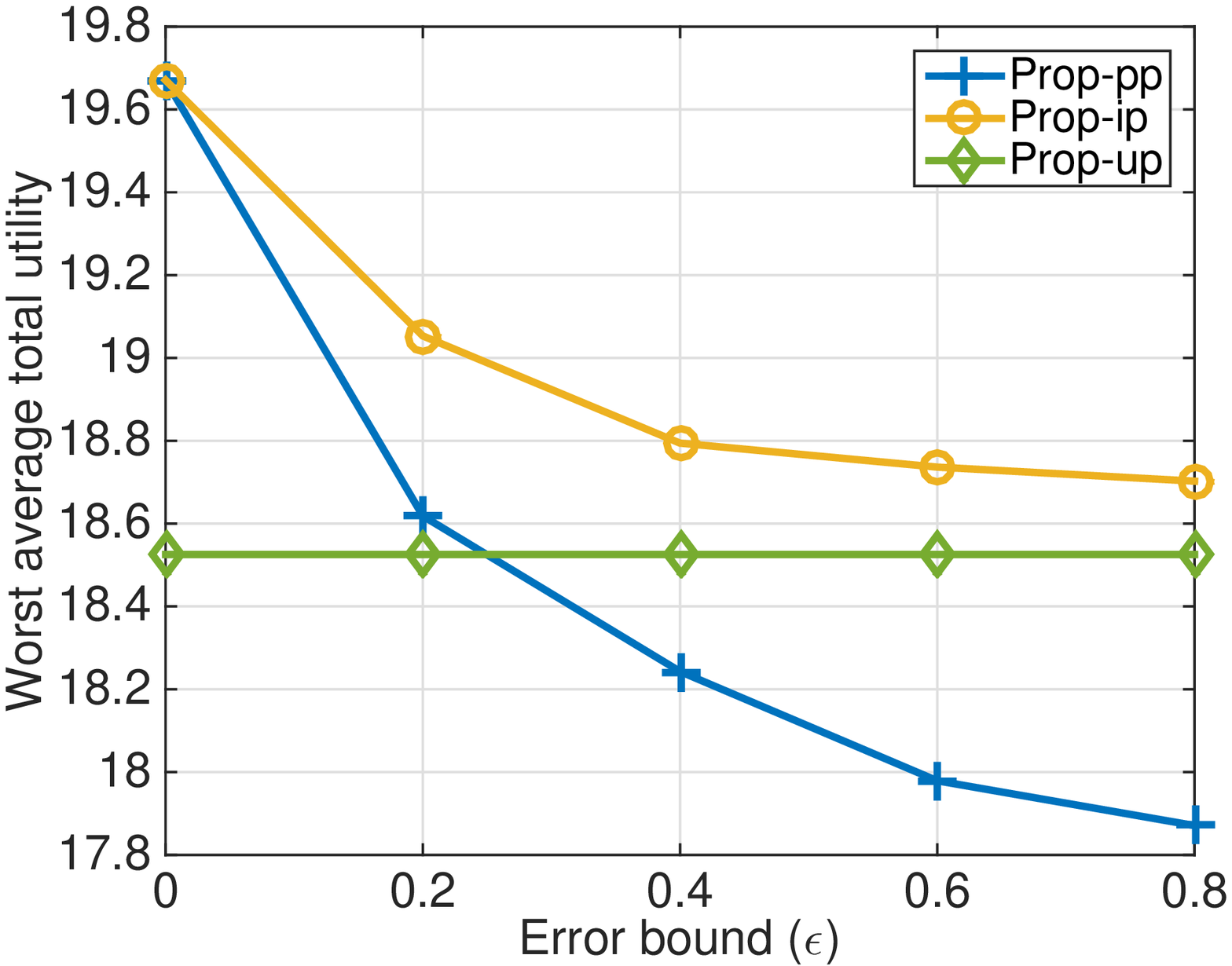}}}
 \end{center}
 \vspace*{-0.37cm}
   \caption{\small{Total utility comparision of the proposed solutions at $M$ = 64.}}
   \label{multi_quality_error}
\vspace*{-0.60cm}
\end{figure}

We consider wireless streaming of the five 360 VR videos given in Table \ref{table} to the five users, respectively. First, we investigate properties of the proposed solutions in the three cases. Fig. \ref{multi_quality_error} (a) illustrates the total utility versus the number of quality levels $L$. Notice that the total utility of Prop-$t$ does not change with $L$, for all $t$ = pp, ip, up. From Fig. \ref{multi_quality_error} (a), we can see that in case-$t$, the gap between the total utilities of Prop-$t$ and Prop-Disc-$t$ decreases with $L$. Furthermore, the gap is small when $L$ is large, implying that the performance loss due to continuous relaxation is negligible at a large $L$. Fig. \ref{multi_quality_error} (b) shows the worst average total utility versus the estimation error bound $\varepsilon$. Note that the worst average total utility of Prop-up is irrelevant to $\varepsilon$. we can see that in the case of imperfect FoV viewing probability distributions, the worst average total utility of Prop-ip is greater than those of Prop-pp and Prop-up, which reveals the importance of explicitly considering imperfectness of the predicted FoV viewing probability distributions in this case; and the worst-case average total utility of Prop-up is greater than that of Prop-pp when $\varepsilon$ is large, as Prop-up is designed to maximize the worst-case total utility and does not depend on any information of FoV viewing probability distributions. Furthermore, the gain of Prop-ip over Prop-pp increases with $\varepsilon$, as it is more important to take into account of FoV prediction error when $\varepsilon$ is large; and the gain of Prop-ip over Prop-up decreases with $\varepsilon$, as the imperfect FoV viewing probability distributions become less important when $\varepsilon$ is large. Fig. \ref{result_multi} illustrates the heatmap of the encoding rates of all tiles for user 4 given by the proposed solutions in the three cases. From Fig. \ref{result_multi} (a) and (b), we can see that the encoding rates of the tiles in an FoV with a larger viewing probability are higher. From Fig. \ref{result_multi} (c), we can tell that the encoding rates of the tiles given by Prop-up are identical. Such observations are in accordance with the optimality properties in Statement (\romannumeral2) of Theorem \ref{theorem}.

\begin{figure}[t]
\begin{center}
 \subfigure[\small{Proposed-pp}]
 {\resizebox{2.8cm}{!}{\includegraphics{pic/result_multi_case1.eps}}}
 \vspace*{-0.07cm}
  \subfigure[\small{Proposed-ip}]
 {\resizebox{2.8cm}{!}{\includegraphics{pic/result_multi_case2.eps}}}
   \subfigure[\small{Proposed-up}]
   {\resizebox{2.8cm}{!}{\includegraphics{pic/result_multi_case3.eps}}}
 \end{center}
 \vspace*{-0.37cm}
   \caption{\small{Encoding rates of the tiles for user 4 given by Prop-pp, Prop-ip and Prop-up at $M$ = 64, $\varepsilon = 0.4$.}}
   \label{result_multi}
\vspace*{-0.2cm}
\end{figure}

\begin{figure}[t]
\begin{center}
 {\resizebox{5.5cm}{!}{\includegraphics{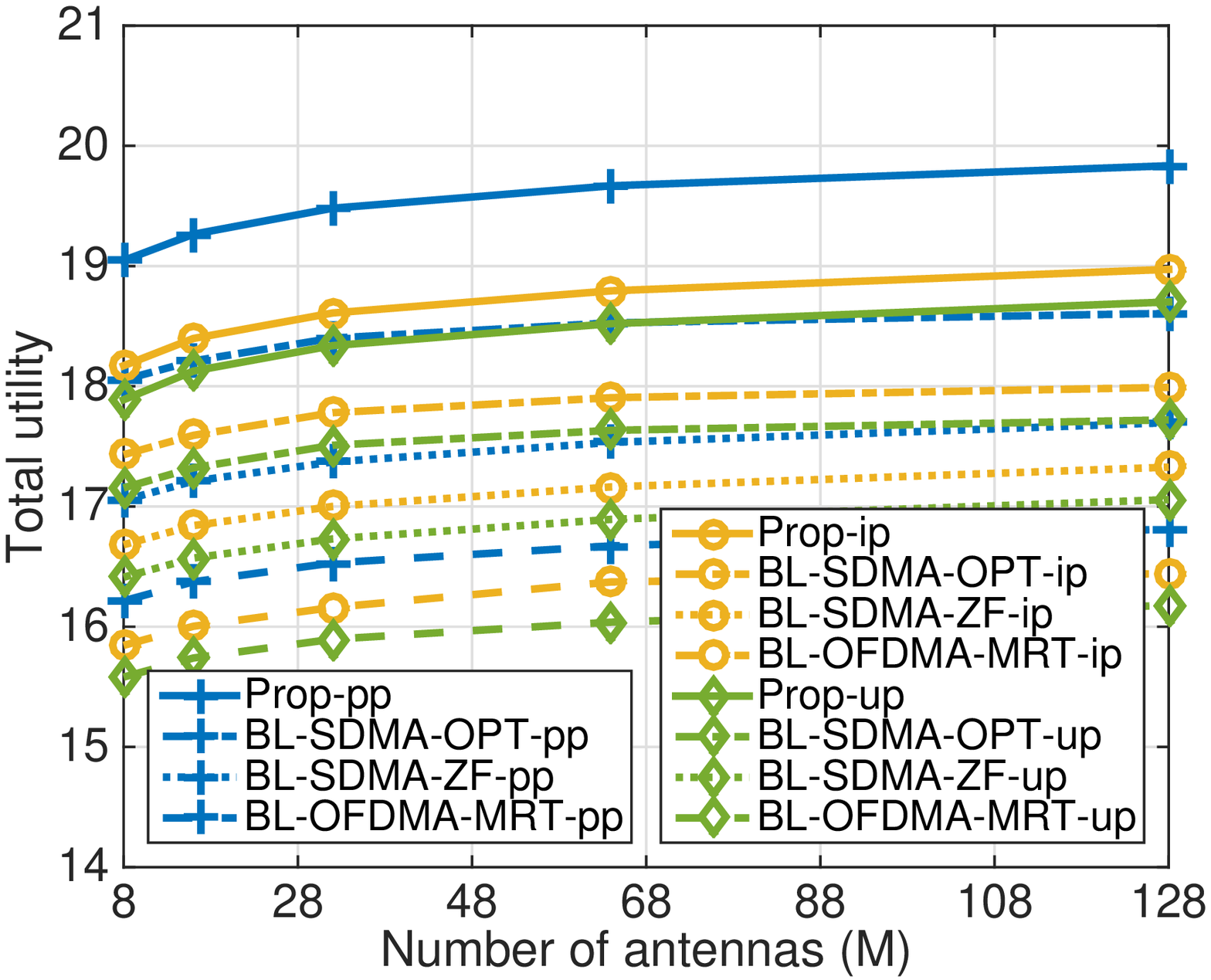}}}
 \vspace*{-0.17cm}
\end{center}
   \caption{\small{Total utility comparision between the proposed solutions and baseline schemes at $\varepsilon = 0.4$.}}
   \label{case_one_utility}
\vspace*{-0.60cm}
\end{figure}

Next, we compare the total utilities of the proposed solution and three baseline schemes, namely BL-SDMA-OPT, BL-SDMA-ZF, BL-OFDMA-MRT, in each case. In BL-SDMA-OPT and BL-SDMA-ZF, SDMA is adopted. In BL-SDMA-OPT, beamforming optimization is considered on each subcarrier, and a suboptimal solution of Problem \ref{rs_case_general} with $d_{c,k} = 0, k\in\mathcal{K}$ and $\mathbf{w}_{c,n} = 0, n\in\mathcal{N}$ (which is a DC programming) is obtained using CCCP. In BL-SDMA-ZF, zero forcing beamforming is adopted on each subcarrier, and an optimal solution of Problem \ref{rs_case_general} with the zero forcing beamformers (which is convex) is obtained by standard convex optimization methods. In BL-OFDMA-MRT, OFDMA is adopted, the maximum ratio transmission (MRT) is adopted for each user, and an optimal solution of Problem \ref{rs_case_general} with the MRT beamformers is obtained by continuous relaxation and the KKT conditions. Fig. \ref{case_one_utility} illustrates the total utility versus the number of transmit antennas $M$. From Fig. \ref{case_one_utility}, we can see that in each case, the proposed solution outperforms the baseline schemes, and the total utility of each scheme increases with $M$. In each case, the gain of the proposed solution over BL-SDMA-OPT arises from the fact that the cost to suppress interference in BL-SDMA-OPT can be high, while the rate splitting scheme partially decodes interference and partially treats interference as noise; the gain of the proposed solution over BL-SDMA-ZF arises from the fact that the cost to suppress interference in BL-SDMA-ZF can be high and the beamforming directions are optimized; and the gain of the proposed solution over BL-OFDMA-MRT is due to effective spatial multiplexing.

Finally, from Fig. \ref{multi_quality_error} (a) and Fig. \ref{case_one_utility}, we can see that $U^{(\text{pp})\dag} > U^{(\text{ip})\dag} > U^{(\text{up})\dag}$, where $U^{(t)\dag}$ represents the total utility of Prop-$t$, where $t\in\{\text{pp},\text{ip},\text{up}\}$. The relationship among the total utilities of the suboptimal solutions of Problem \ref{rs_case_general} in the three cases is the same as that of the optimal solutions in the three cases (which is shown in Statement (\romannumeral3) of Theorem \ref{theorem}).


\section{Conclusion}
In this paper, we investigated optimal designs, based on rate splitting with successive decoding, for streaming multi-quality tiled 360 VR videos from a multi-antenna server to multiple single-antenna users in a multi-carrier system. We considered three cases of FoV viewing probability distributions and proposed three respective performance metrics for total utility. In each case, we optimized the rate adaptation, resource allocation and beamforming to maximize the total utility. We successfully transformed the challenging nonconvex problem in each case into a DC programming problem, and obtained a suboptimal solution using CCCP. Finally, numerical results demonstrated the proposed solutions achieve notable gains over existing schemes in all three cases, and revealed the impact of FoV prediction and its accuracy on streaming of multi-quality tiled 360 VR videos.


\end{document}